f
\documentclass[showpacs,preprintnumbers,preprint,aps,prd,amsmath,amssymb]{revtex4-1}
\pdfoutput=1

\usepackage{amssymb}
\usepackage{mathrsfs}
\usepackage{slashed}
\usepackage{graphicx,color}
\usepackage{amsmath}
\usepackage{subeqnarray}
\usepackage{multirow}
\allowdisplaybreaks

\def\be{\begin{equation}}
\def\ee{\end{equation}}
\def\bea{\begin{eqnarray}}
\def\ba{\begin{array}}
\def\ea{\end{array}}
\def\eea{\end{eqnarray}}\def\nn{\nonumber}
\def\bse{\begin{subequations}}
\def\ese{\end{subequations}}

\begin{document}
\author{Shivani Gupta \footnote{shivani@yonsei.ac.kr}}
\affiliation{Department of Physics and IPAP, Yonsei University, Seoul 120-749, Korea}
\author{C. S. Kim \footnote{cskim@yonsei.ac.kr}}
\affiliation{Department of Physics and IPAP, Yonsei University, Seoul 120-749, Korea}
\author{Pankaj Sharma \footnote{pankajs@kias.re.kr}}
\affiliation{Korea Institute of Advanced Study, Seoul 130-722, Korea}
\vspace*{1cm}

\title{Radiative and Seesaw Threshold Corrections \\
to the $S_3$ Symmetric Neutrino Mass Matrix}

\begin{abstract}
We systematically analyze the radiative corrections to the $S_3$ symmetric neutrino mass matrix
at high energy scale, say the GUT scale, in the charged lepton basis.
There are significant corrections to the neutrino parameters both in the Standard Model (SM) and
Minimal Supersymmetric Standard Model (MSSM) with large
tan$\beta$, when the renormalization group evolution (RGE) and seesaw threshold effects are taken into consideration.
We find that in the SM all three mixing angles and atmospheric mass squared
difference are simultaneously obtained in their current 3$\sigma$ ranges at the electroweak scale.
However, the solar mass squared difference is found to be larger than its allowed 3$\sigma$ range at the low scale
in this case. There are significant contributions to neutrino masses and mixing angles in the MSSM with large tan$\beta$
from the RGEs even in the absence of seesaw threshold corrections.
However, we find that the mass squared differences and the mixing angles are obtained in their current 3$\sigma$
ranges at low energy when the seesaw threshold effects are also taken into account in the MSSM with large tan$\beta$.

\end{abstract}

\maketitle

\section{Introduction}

The neutrino oscillation experiments have enriched our knowledge of masses and mixings of neutrinos
and thus flavor structure of leptons. These developments aspire theorists to construct models
for unraveling the symmetries of lepton mass matrices.
For three neutrino flavors the solar and atmospheric mass squared differences are given as,
$\Delta m_{12}^2 \approx 10^{-5}$ eV$^2$ and $|\Delta m_{13}|^2 \approx 10^{-3}$ eV$^2$, where $\Delta m_{ij}^2 =m_j^2-m_i^2$
and $m_{i,j}$ are the mass eigenvalues of neutrinos, respectively. With the evidence of nonzero
value of reactor mixing angle $\theta_{13}$ \cite{theta13} we now have information of all three mixing angles
contrary to earlier studies where only an upper bound on $\theta_{13}$ existed.
The lepton flavor mixing matrix comprising of three mixing angles and a CP violating phase is given as
\be
U=
\left(
\ba{ccc}
c_{12}c_{13} & s_{12}c_{13} & s_{13}e^{-i\delta_{CP}} \\
-s_{12}c_{23}-c_{12}s_{23}s_{13}e^{i\delta_{CP}} &
c_{12}c_{23}-s_{12}s_{23}s_{13}e^{i\delta_{CP}} & s_{23}c_{13} \\
s_{12}s_{23}-c_{12}c_{23}s_{13}e^{i\delta_{CP}} &
-c_{12}s_{23}-s_{12}c_{23}s_{13}e^{i\delta_{CP}} & c_{23}c_{13}
\ea
\right).
\ee
Here $c_{ij}=\cos\theta_{ij}, s_{ij}=\sin\theta_{ij}$; $\theta_{ij}$ are the three mixing angles,
$\delta_{CP}$ is the Dirac CP phase. There are two additional CP phases if neutrinos are Majorana particles.
The best fit values along with their 3$\sigma$ ranges of neutrino oscillation parameters \cite{fogli} are shown in Table I.
Dirac phase $\delta_{CP}$, which contributes to CP violation in the leptonic sector, is expected to be measured
in the long baseline neutrino oscillation experiments. The strength
of this leptonic CP violation is parameterized  by Jarlskog rephasing invariant \cite{jarlskog} $J=c_{12}s_{12}c_{23}s_{23}c_{13}^2s_{13} \sin \delta_{CP}$.
The two Majorana phases, however, contribute to the lepton number violating processes like neutrinoless double beta decay.
The possible measurement of effective Majorana mass in the current and upcoming experiments like GERDA, EXO, CUORE, MAJORANA, SuperNEMO \cite{NDBD}
will provide some additional constraints on the two Majorana phases and neutrino mass scale.
The cosmological constraint on the sum of neutrino masses by the Planck Collaboration \cite{Planck} is
$\Sigma m_{\nu_i} < $0.23 eV at 95$\%$ C.L.. Depending on the values chosen for the priors this sum can be in the range $(0.23-0.933)$ eV.
An additional bound from the Solar Pole Telescope Collaboration \cite{Hou} gives $\Sigma m_{\nu_i} =$ 0.32$\pm$ 0.11 eV.
However, these bounds are expected to be tested in forthcoming observations.
There are some challenges left namely to determine the absolute mass scale, mass hierarchy of neutrinos
and the CP violation in leptonic sector amongst others.

\begin{table}[t]
\begin{center}
\begin{tabular}{lccccc}
\hline
\hline
Parameter & ~~~~~~~~ & Best fit  & ~~~~~~~~ & $3\sigma$ \\
\hline
$\Delta m^2_{12}/10^{-5}~\mathrm{eV}^2 $ (NH or IH) & & 7.54 & & 6.99 -- 8.18 \\
\hline
$\Delta m^2_{13}/10^{-3}~\mathrm{eV}^2 $ (NH) & & 2.43 & & 2.19 -- 2.62\\
\hline
$\theta_{12}$ & & 33.64$^\circ$  & & 30.59$^\circ$  -- 36.8$^\circ$  \\
\hline
$\theta_{13}$ & & 8.93$^\circ$ & & 7.47$^\circ$ --10.19$^\circ$  \\
\hline
$\theta_{23}$ & & 37.34$^\circ$ & & 35.1$^\circ$ -- 52.95$^\circ$                   \\
\hline
\label{Table:parameters}
\end{tabular}
\begin{center}
\caption{The experimental constraints on neutrino parameters taken from \cite{fogli}.}
 \end{center}
 \end{center}
\end{table}

Evidence for nonzero $\theta_{13}$ has lead to many
studies for the deviation from the assumed  symmetries that predict the vanishing $\theta_{13}$ value. Among many possible
discrete flavor symmetries to produce the current data, $S_3$ has been extensively studied in literature \cite{S3}.
It is the smallest discrete non-Abelian group which is the permutation of three objects. Perturbations to
$S_3$ symmetric leptonic mass matrices have been used to study the mass spectra of the leptons and predict
well known democratic \cite{democratic} and tri-bimaximal neutrino mixing scenario \cite{TBM}.
The effective light neutrino mass matrix, M$_{\nu}$, invariant under $S_3$ is given as
\be
\label{Mnu}
M_{\nu}=pI + qD,
\ee
where
\be
I=
\left(
\ba{ccc}
1 &0 &0 \\
0 & 1 &0 \\
0 & 0 & 1
\ea \right), ~~~~~
D=\left(
\ba{ccc}
1&1&1 \\
1&1&1 \\
1&1&1
\ea \right).
\ee
Here $D$ is democratic matrix, $p$ and $q$ are in general complex parameters.
There are quite a few studies about the breaking of $S_3$ symmetry in the leptonic sector \cite{Radhe, zhou, Jora}
to produce the current observed neutrino oscillation data.
It will be interesting to study order of breaking
in the $S_3$ symmetric neutrino mass matrix that arise due to the RGE
and the seesaw threshold corrections between the GUT scale $\Lambda_g$ and the electroweak scale $\Lambda_{ew}$ in the SM
and MSSM.
In the light of non zero $\theta_{13}$ our main aim is to study the viability
of producing the current neutrino oscillation data at $\Lambda_{ew}$ scale in an $S_3$ symmetric
neutrino mass matrix $M_{\nu}$ at $\Lambda_g$ when both radiative and seesaw threshold corrections are taken into account.

Earlier studies have shown that there are significant RGE corrections to
neutrino masses and mixing angles particularly for quasi-degenerate neutrino spectrum in the MSSM with large tan$\beta$.
The SM is extended by three heavy right handed neutrinos at high energy scale to generate neutrino masses in
Type I seesaw mechanism \cite{seesaw}. The seesaw threshold corrections arise due to subsequent decoupling
of these heavy right handed Majorana neutrinos at their respective masses.
The structure of the Dirac mass matrix $M_D$ is proportional to the neutrino Yukawa coupling matrix $Y_{\nu}$.
We take a general $Y_{\nu}$ and scan the parameter space to obtain the desired mixing pattern.
The right handed Majorana mass matrix $M$ is found by inverting the Type I seesaw formula at $\Lambda_g$.
The charged lepton mass matrix $M_l$ at $\Lambda_{g}$ is taken to be diagonal.
We study the running behavior of neutrino masses and mixings from $\Lambda_{g}$ down
to $\Lambda_{ew}$ using the RGE
for the Yukawa couplings in the $S_3$ symmetric $M_{\nu}$ both in the SM and MSSM.
Above the heaviest seesaw scale ($M_3$) there is a full theory and thus
RGEs for Yukawa couplings $Y_e$, $Y_{\nu}$ and mass matrix $M$ are considered.
However, since our right handed neutrino mass matrix $M$ is hierarchical ($M_1<M_2<M_3$),
we also consider the seesaw threshold effects and thus the respective set of effective theories in between these scales,
arising from the subsequent decoupling of heavy right handed fields at their respective masses.
In the SM we find that the atmospheric mass squared difference
($\Delta m^2_{13}$) along with the neutrino mixing angles are generated in their present 3$\sigma$ ranges at the low energy scale.
However, the solar mass squared difference ($\Delta m^2_{12}$) is greater than its allowed value ($\approx 10^{-4}$) in the SM.
We find that it is possible to radiatively generate the current neutrino masses and mixing angles
from the $S_3$ invariant neutrino mass matrix $M_\nu$ in the charged lepton basis,
when the seesaw threshold effects are taken into account in the MSSM with large tan$\beta$.
In the MSSM with large tan$\beta$ the solar mass squared difference $\Delta m^2_{12}$
can be produced in its current range along with the other neutrino oscillation parameters at $\Lambda_{ew}$
in the presence of these threshold corrections.

In section II we give the form of lepton mass matrices considered at $\Lambda_g$.
In the subsequent section, we give the RGE equations governing from $\Lambda_g$ to $\Lambda_{ew}$,
in presence of the seesaw threshold effects both in the SM and MSSM. In section IV we study
the order of corrections to the neutrino mass matrix in the presence of seesaw threshold effects.
Section V gives our numerical results for both cases under consideration.
We conclude in the last section.

\section{Form of lepton mass matrices at the GUT scale}

We consider the basis where charged lepton mass matrix $(M_l)$ is diagonal and the effective light neutrino
mass matrix $(M_{\nu})$ is $S_3$ symmetric as given in Eq. (\ref{Mnu}).
The Yukawa coupling matrix for charged leptons is given as
\be
Y_e =\frac{1}{v}\left(
\ba{ccc}
m_e &0 &0 \\
0 & m_{\mu} &0 \\
0 & 0 & m_{\tau}
\ea \right),
\ee
where the Higgs vacuum expectation value (VEV) $v$ is taken to be 246 GeV in the SM and 246$\cdot\cos\beta$ GeV in the MSSM.
The Yukawa coupling matrix $Y_{\nu}$ for the light neutrinos is taken of the form $Y_{\nu}= y_{\nu} U_{\nu} D$ as given in \cite{Xing}
where $D$ is the diagonal matrix Diag$(r_1,r_2,1)$. The three parameters $y_\nu$, $r_1$ and $r_2$ are real,
positive dimensionless parameters that characterize eigenvalues of $Y_{\nu}$.
The unitary matrix $U_{\nu}$ is the product of the three rotation
matrices $R_{23}(\vartheta_2)\cdot R_{13}(\vartheta_3 e^{-i\delta})\cdot R_{12}(\vartheta_{1})$ having one CP violating phase $\delta$.
Thus, $Y_{\nu}$ has seven unknown parameters viz. three eigenvalues, three mixing angles and one CP phase.
We vary the three hierarchy ($y_{\nu}$, $r_1$, $r_2$) parameters and, though they are completely arbitrary,
but assumed to be $< \mathcal{O}(1)$. Three angles $\vartheta_1$, $\vartheta_2$, $\vartheta_3$ and $\delta$
are varied in the range of (0-2$\pi$).

The right handed mass matrix $M$ is found by inverting the Type I seesaw formula as
\be
\label{MR}
M= -Y_{\nu} M_{\nu}^{-1}Y_{\nu}^T,
\ee
$i.e.$ $M$ in our study is found from $Y_{\nu}$ and the $S_3$ symmetric neutrino mass matrix $M_{\nu}$ at $\Lambda_g$ by inverting seesaw formula.
The three right handed neutrino masses $M_1$, $M_2$ and $M_3$
are obtained by diagonalizing the right handed Majorana mass matrix $M$.
The light neutrino mass matrix $M_{\nu}$ can be diagonalized by the unitary transformation $R$ as $R^T M_{\nu} R$. One of the possible form of R
can be
\be
R = U_{TBM} =\left(
\ba{ccc}
\frac{2}{\sqrt{6}} &\frac{1}{\sqrt{3}} & 0 \\
-\frac{1}{\sqrt{6}}  & \frac{1}{\sqrt{3}}  &-\frac{1}{\sqrt{2}} \\
-\frac{1}{\sqrt{6}}  & \frac{1}{\sqrt{3}}  & \frac{1}{\sqrt{2}}
\ea \right) .
\ee
The mass eigenvalues of $M_{\nu}$ are $p$, $p+3q$ and $p$: corresponding to the light neutrino masses $m_1$, $m_2$ and $m_3$, respectively.
Due to the degeneracy in the mass eigenvalues $m_1$ and $m_3$,
the diagonalizing matrix $R$ is not unique. Degeneracy of masses implies that $R$ is arbitrary
up to orthogonal transformation $R_{13}(\phi)$, where $\phi$ is in 1-3 plane. Thus, most general diagonalizing matrix R
is $U_{TBM}R_{13}(\phi)$, which implies the same physics as $U_{TBM}$. In this work we set $\phi$=0 without loss of generality \cite{ Radhe, Jora}.
From the neutrino oscillation data we know $\Delta m_{12}^2 \approx$10$^{-5}$ and thus, there is small difference in the mass eigenvalues
$m_1$ and $m_2$ which is a possible objection to this scenario of $S_3$ invariant
approximation as here $m_1$ and $m_3$ are degenerate. The possible solution to this problem is given in \cite{Radhe, Jora}
where the complex numbers $p$ and $p+3q$ are considered to have same magnitude but different directions.
As shown in \cite{Radhe}, $q$ can be chosen completely imaginary and $p$ is taken to be $|p|e^{-i\frac{\alpha}{2}}$.
The magnitudes of $p$ and $q$ can be written in terms of parameter $x$ as
\bea \nn
\label{magnitude}
|p| = x\ sec\frac{\alpha}{2}, \\
|q| = \frac{2}{3}x\ tan\frac{\alpha}{2},
\eea
where $x$ is a real free parameter and allowed range of $\alpha$ is $0\leq\alpha<\pi$.
The magnitude of $p$ and $p+3q$ can be made equal by adjusting the phase $\alpha$. The parameter $x$ vanishes when $\alpha$=180$^\circ$
and thus this value is disallowed. Substituting the values of $p$ and $q$ given in Eq. (\ref{magnitude}),
the magnitudes of the mass eigenvalues are given as
 \be
 |m_1| =|m_2|=|m_3|= x\ sec\frac{\alpha}{2}.
 \ee
This results in equal magnitude of all three mass eigenvalues and thus a degenerate spectrum of neutrinos to begin with at $\Lambda_g$.
As pointed out earlier in \cite{Radhe} that the phase $\alpha$ affects the rate of neutrinoless double beta decay
but will not affect neutrino oscillation parameters. Thus, this phase is of Majorana type.
When we run these masses from $\Lambda_{g}$ to highest seesaw scale $M_3$, the degeneracy of the mass eigenvalues is lifted by RGE corrections.
We consider the normal hierarchical spectrum of masses where $m_1$ is the lowest mass. The other two masses are given as
$m_2=\sqrt{m_1^2+\Delta m_{12}^2}$ and $m_3=\sqrt{m_1^2+\Delta m_{13}^2}$. Since the three mass eigenvalues at $\Lambda_g$
have equal magnitude the two mass squared differences are vanishing to begin with.
Once the degeneracy of three mass eigenvalues is lifted, their non zero values are generated.
In subsequent sections we will explore the generation of the solar and atmospheric mass squared differences, together with the three mixing angles
in their current 3$\sigma$ limit at $\Lambda_{ew}$ through the radiative corrections
from $S_3$ invariant neutrino mass matrix at $\Lambda_{g}$.

\section{RGE equations with seesaw threshold effects}

In Type I seesaw the SM is extended by introducing three heavy right handed neutrinos and keeping the Lagrangian of
electroweak interactions invariant under $SU(2)_L\times U(1)_Y$ gauge transformation.
In this case, the leptonic Yukawa terms of the Lagrangian are written as
\be \label{typei-lag}
-\mathcal L_{\nu} = \bar l_e \phi Y_e e_R +\bar l_e \tilde \phi Y_{\nu} \nu_R +\frac{1}{2} \bar {\nu_R^{c}} M\nu_R + h.c..
\ee
Here $\phi$ is the SM Higgs doublet and $\tilde \phi= i\sigma^2 \phi^{*}$. $e_R$ and $\nu_R$
are right handed charged lepton and neutrino singlets, $Y_e$ and $Y_\nu$ are the Yukawa coupling matrices for
charged leptons and Dirac neutrinos, respectively. The last term of Eq. (\ref{typei-lag}) is the Majorana mass
term for the right handed neutrinos.

Quite intensive studies have been done in literatures \cite{Antush, Antush2} regarding the general features of RGE of neutrino parameters.
At the energy scale below seesaw threshold $i.e.$ when all the heavy particles are integrated out,
the RGE of neutrino masses and mixing angles is described by the effective theory which is same
for various seesaw models. But above the seesaw scale full theory has to be considered
and thus, there can be significant RGE effects due to the interplay of heavy and light sector.
The RGE equations and subsequent decoupling of heavy fields at their respective scales are elegantly
given in Refs. \cite{seesawcorrections, ellis, jianwei}. The comprehensive study of the RGE and seesaw threshold corrections
to various mixing scenarios is recently done in \cite{ourpap}.

The effective neutrino mass matrix $M_{\nu}$ above $M_3$ is given as
\be
M_{\nu}(\mu) =-\frac{v^2}{2}Y_{\nu}^T(\mu) M^{-1}(\mu) Y_{\nu} (\mu),
\ee
where $v$ = 246$\cdot\sin\beta$ GeV in the MSSM and $\mu$ is the renormalization scale. $Y_{\nu}$ and $M$ are $\mu$ dependent.
Since we study the evolution of leptonic mixing parameters from $\Lambda_{g}$ to $\Lambda_{ew}$ scale in a generic seesaw model
we need to take care of the series of effective theories that arise by subsequent decoupling of the heavy right handed fields $M_i$ ($i=$1,2,3)
at their respective mass thresholds.
The Yukawa couplings $Y_{\nu}$ and $M$ are dependent on the energy scale $\Lambda$.
At the GUT scale we consider the full theory and the one loop RGEs for $Y_e$, $Y_{\nu}$ and $M$ are given as
\bea \nn
 \Dot{Y_e}&=& \frac{1}{16 \pi^2}Y_{e}\left[\alpha_e+C_1 H_e + C_2 H_{\nu}\right],\nn \\
 \Dot{Y_{\nu}}&= \nn& \frac{1}{16 \pi^2}Y_{\nu}\left[\alpha_{\nu}+C_3 H_e + C_4 H_{\nu}\right],\\
 \Dot{M}&= &\frac{1}{16 \pi^2}C_5\left[(Y_{\nu}Y_{\nu}^{\dagger})M + M (Y_{\nu}Y_{\nu}^{\dagger})^T\right],
\eea
where $\Dot{Y_i}=\frac{dY_i}{dt}$ ($i$=$e, \nu$), t$=ln(\mu/\mu_0)$ with $\mu$($\mu_0$)
being the running (fixed) scale, and $H_i = Y_{i}^{\dagger} Y_i$ ($i= e, \nu$).
The coefficients are $C_1=\frac{3}{2}, C_2=-\frac{3}{2}, C_3=-\frac{3}{2}, C_4= \frac{3}{2}, C_5=1$ in the SM and
 $C_1=3, C_2=1, C_3=1, C_4= 3, C_5=2$ in the MSSM, respectively. The expressions for
$\alpha_e$ and $\alpha_{\nu}$ in the SM and MSSM are explicitly given as
\bea \nn
\label{alphamssm}
\alpha_{e(SM)} &=& Tr(3H_u +3H_d+ H_e+H_\nu)-(\frac{9}{4}g_1^2 + \frac{9}{4}g_2^2),\\ \nn
\alpha_{\nu(SM)} &=& Tr(3H_u +3H_d+ H_e+H_\nu)-(\frac{9}{20}g_1^2 + \frac{9}{4}g_2^2), \\\nn
\alpha_{e(MSSM)} &=& Tr(3H_d+H_e)-(\frac{9}{5}g_1^2+3g_2^2),\\
\alpha_{\nu(MSSM)} &=& Tr(3H_u+H_{\nu})-(\frac{3}{5}g_1^2+3g_2^2),
\eea
where $g_{1,2}$  are the $U(1)_Y$ and $SU(2)_L$ gauge coupling constants.

The heavy right handed mass matrix $M$ obtained from Eq. (\ref{MR}) is non diagonal
and thus is diagonalized by the unitary transformation $U_R$ as
\be
\label{MR2}
U_R^T M U_R = Diag{(M_1,M_2,M_3)}.
\ee
The Yukawa coupling $Y_{\nu}$ is accordingly transformed as $Y_{\nu}U_R^*$.
At the highest seesaw scale $M_3$, the effective operator $\kappa_{(3)}$ is given by the matching condition as
\be
\label{kappa3}
\kappa_{(3)}= 2 Y_{\nu}^T M_3^{-1}Y_{\nu},
\ee
in the basis where $M$ is diagonal. The Yukawa coupling $Y_{\nu}$ above is a 3$\times$3 matrix and all the variables are set to the scale $M_3$.
At the scale lower than $M_3$ ($\mu < M_3$) the effective neutrino mass matrix $M_{\nu}$ is given as
\be
\label{Mnuthr}
M_{\nu} = -\frac{v^2}{4}\{ \kappa_{(3)} + 2 Y_{\nu(3)}^T M_{(3)}^{-1}Y_{\nu(3)}\}.
\ee
As can be seen that  $M_{\nu}$ is the sum of $\kappa_{(3)}$ given in Eq. (\ref{kappa3}) and the seesaw factor which is obtained
after decoupling $M_3$. Thus, $Y_{\nu (3)}$ is $2 \times 3$ and $M_{(3)}$ is $2 \times 2$ mass matrices.
RGE between the scale $M_3$ and $M_2$ is governed by the running of $\kappa_{(3)}$, $Y_{\nu(3)}$ and $M_{(3)}$.
The running of $\kappa_{(3)}$ is given as
\bea
\label{running}
\Dot{\kappa}_{(3)(SM)}&= &\frac{1}{16 \pi^2} \left[(C_3 H_e + C_6 H_{\nu (3)})^T \kappa_{(3)} +
\kappa_{(3)} (C_3 H_e + C_6 H_{\nu (3)}) + \alpha_{(3)} \kappa_{(3)}\right], \\ \nn
\Dot{\kappa}_{(3)(MSSM)}&=& \frac{1}{16 \pi^2}\left[(H_e+C_6 H_{\nu (3)})^T\kappa_{(3)}+\kappa_{(3)}(H_e +C_6 H_{\nu (3)}) + \alpha_{(3)} \kappa_{(3)}\right],
\eea
where $C_6 = \frac{1}{2}$ and $H_{\nu (3)} = Y_{\nu (3)}^{\dagger}Y_{\nu (3)}$. $\alpha_{(3)}$ in the SM and MSSM is explicitily given as
\bea \nn
\alpha_{(3)(SM)} & = & 2\ Tr (3H_d + 3H_u + H_e + H _{\nu (3)})- 3g_2^2+\lambda, \\ \nn
\alpha_{(3)(MSSM)} &=& 2\ Tr (3H_u+ H_{\nu (3)})-\frac{6}{5}g_1^2-6g_2^2.
\eea
The effective operator $\kappa_{(2)}$ at $M_2$ is given by the matching condition as
\be
\kappa_{(2)} = \kappa_{(3)} + 2 Y_{\nu (2)}^T M_{(2)}^{-1} Y_{\nu(2)}.
\ee
where all the parameters are set to seesaw scale $M_{2}$.
 $M_{\nu}$ at the scale below $M_2$ has the RGE equation given in Eq. (\ref{running}). In their RGE expression
we have $Y_{\nu (2)}$ which is $1 \times 3$ and $M_{(2)}$ which is $1 \times 1$.
The low energy effective theory operator $\kappa_{(1)}$ is obtained after integrating out all three heavy right handed fields.
The one loop RGE for $\kappa_{(1)}$ from lowest seesaw scale $M_1$ down to $\Lambda_e$ scale is given as
\be
 \Dot{\kappa}_{(1)}=  (C_3 H_e^T) \kappa_{(1)} + \kappa_{(1)}(C_3 H_e) +\alpha \kappa_{(1)},
\ee
where
\be
\label{alphasm}
\alpha= 2\ Tr(3H_u+3H_d + H_e)  - 3g_2^2 +\lambda \mbox{~~in the SM},
\ee
\be \nn
\alpha= 2\ Tr(3H_u)-\frac{6}{5}g_1^2-6g_2^2 \mbox{~~in the MSSM.}
\ee
When the Higgs field gets VEV, the light neutrino mass matrix is obtained from $\kappa_{(1)}$
as $M_{\nu}=\frac{\kappa_{(1)}v^2}{4}$. We diagonalize $M_\nu$ to obtain neutrino masses, mixing angles and CP phases.

\section{Neutrino masses and mixings}

The RGE above the highest seesaw scale $M_3$ depends on more parameters than below
the lowest seesaw scale $M_1$ due to presence of the neutrino Yukawa couplings $(Y_{\nu})$.
The RGE equations consist of $H_e$, $M$ and $H_{\nu}$ out of which latter can be large.
In the basis where charged lepton mass matrix is diagonal, $M_{\nu}$ at two different energy scales
$\Lambda_{ew}$ and $\Lambda_{g}$ are homogeneously related as \cite{ellis, sruba, dighe}
\be
M_{\nu}^{\Lambda_{ew}} = I_K\cdot I^T\cdot M_{\nu}^{\Lambda_{g}}\cdot I.
\ee
Here $I_K$ is flavor independent factor arising from gauge interactions and fermion antifermion loops. It does not influence the mixing angles.
The matrix $I$ has the form
\bea \nn
I&=& Diag (e^{-\Delta_e},e^{-\Delta_{\mu}},e^{-\Delta_{\tau}}), \\
 &\backsimeq& Diag(1-\Delta_e,1-\Delta_{\mu},1-\Delta_{\tau}) +\mathcal{O}(\Delta_{e,{\mu},{\tau}}^2),
\eea
where
\be
\label{rgmag}
\Delta_j =\frac{1}{16\pi^2}\int[3(H_j)-(H_{\nu_j})]dt,
\ee
where $j=e, {\mu}, {\tau}$.
Numerically, $\Delta^{SM}_{\tau}$ can be of the order of 10$^{-3}$ when $Y_{\tau} \sim .01$ and $Y_{\nu}=0.2$ and the scale $\mu$ and $\mu_0$
are 10$^{12}$ and 10$^2$, respectively. In the MSSM, $\Delta^{MSSM}_{\tau}$ $\sim$10$^{-3} (1+\tan^2 \beta)$ for the same values
of $Y_{\nu}$ and $Y_{\tau}$. In the absence of seesaw threshold effects, $\Delta_{\tau}$
is small $\approx$ 10$^{-5}$ in the SM for the above mentioned values of
 $Y_{\tau}$.

At above seesaw scale appreciable deviations may occur only for large values of $Y_{\tau}$ or $Y_{\nu}$.
In the absence of seesaw threshold effects $i.e.$ when there is no $H_{\nu}$ term in Eq. (\ref{rgmag}) the radiative corrections
are governed by $\Delta_{\tau}$ term as $\Delta_{e,\mu}$ is too small.
On the other hand, due to the presence of large $H_{\nu}$, the seesaw threshold effects play a crucial role in the running of mixing angles.
Below the seesaw scales, deviations are obtained from $Y_{\tau}\sim \sqrt{2}m_{\tau}/v$ $\approx \mathcal{O}(10^{-2})$ in the SM,
and $\sim \frac{\sqrt{2}m_{\tau}}{v \cos \beta}$ in the MSSM. There can be significant deviations in the MSSM with large $\tan\beta$
which enhances $Y_{\tau}$. In the presence of seesaw threshold,
appreciable deviations are possible as it is natural
to have $Y_{\nu} \sim \mathcal{O} (1)$.
The analytic expressions for the running neutrino mixing angles, masses and CP phases are quite
long and have been earlier derived in literatures \cite{jianwei, sruba, bergstrom}.
As shown earlier in \cite{Antush2, Casas}, the mass squared differences in the denominator
of the RGEs play a significant role
in the evolution of neutrino mixing angles and phases, if the left-handed neutrinos are nearly degenerate.

\begin{table}
\centering
\label{results}
\begin{tabular}{c|cccc}
\hline
\hline
Parameters & SM &~~~~~~~~  & MSSM  \\
Input & &~~~~~~~~ &\\
\hline
$r_1$ &7$\times 10^{-3} $ & &$3\times 10^{-4}$\\
$r_2$ & 0.33 & & 0.34\\
$\delta$ & 158.1$^\circ $ & &96.3$\circ$ \\
$y_{\nu}$& 0.72 & &0.49\\
$\theta_1$ & 225.2$^\circ$  & & 188.5$^\circ$  \\
$\theta_2$ & 244.6$^\circ$ & & 245.8$^\circ$  \\
$\theta_3$ & 345.5$^\circ$ & & 241.7$^\circ$  \\
$x (eV)$ & 7.46$\times 10^{-2}$ & & $1.9\times 10^{-3}$  \\
$\alpha$ & 102.5$^\circ$  & & 170.2$^\circ$  \\
\hline
Output & &~~~~~~~~ &\\
\hline
$m_1 (eV)$  & $9.4\times10^{-2}$  & & $1.47\times 10^{-2}$\\
$\theta_{12}$ &34.3$^\circ$ & & 35.2$^\circ$ \\
$\theta_{13}$ &7.88$^\circ$ & & 9.85$^\circ$\\
$\theta_{23}$ &47.7$^\circ$ & & 47.4$^\circ$\\
$\Delta m^2_{12}(eV^2)$ & 4.18 $\times 10^{-4}$& & $7.14\times 10^{-5}$\\
$\Delta m^2_{13}(eV^2)$ & 2.48 $\times 10^{-3}$& & $2.35\times 10^{-3}$\\
$M_{1}$ (GeV)& 3.1 $\times 10^{5}$& & $2.86\times 10^{3}$\\
$M_{2}$ (GeV)& 4.43 $\times 10^{8}$& & $9\times 10^{8}$\\
$M_{3}$ (GeV)& 2.83 $\times 10^{9}$ & & $4.7\times 10^{9}$\\
$J$&-2.99$\times10^{-2}$& &$-3.87\times 10^{-2}$\\
$|m_{ee}|(eV)$&9.2$\times10^{-2}$& &$1.03\times 10^{-2}$\\
\hline
\end{tabular}
\begin{center}
\caption{Numerical values of input and output parameters that are radiatively
generated via the RGE and seesaw threshold effects both in the SM and MSSM.
The input parameters are taken at $\Lambda_{g}$ = 2$\times 10^{16}$ GeV and tan$\beta$=55 in the MSSM.}
 \end{center}
\end{table}

\section{Numerical Results and Discussions}

At the GUT scale $\Lambda_g$ we have seven free parameters in $Y_{\nu}$ and two free parameters $\alpha$ and $x$ in $M_{\nu}$.
The three mixing angles in $\vartheta_1$, $\vartheta_2$ and $\vartheta_3$ and phase $\delta$ are allowed to take the values in the
range (0--2$\pi$). The hierarchy parameters $y_{\nu}$, $r_1$, $r_2$ and $x$ are randomly varied and are expected to be $<\mathcal{O}(1)$.
The physical range of phase $\alpha$ is from (0--$\pi$).
The mass spectrum at the high scale is degenerate and thus we have vanishing solar and atmospheric mass squared difference to begin with.
The parameter space at the $\Lambda_g$ with which the low energy neutrino data is obtained at the $\Lambda_{ew}$ is illustrated in Table II.
The set of input parameters in that particular parameter space are also given in the table.
We choose the set of parameters in parameter space at the high scale for which maximum value of $\theta_{13}$
is obtained and the other mixing angles and mass squared differences are simultaneously obtained in current 3$\sigma$ range at $\Lambda_{ew}$.
However, the parameter space under consideration is only for illustration and not unique.
Search for complete parameter space is an elaborate study and thus independent future work.

\subsection{Radiative Threshold corrections in the SM}

\begin{figure}[b]
\label{figsm}
\begin{center}
\includegraphics[width=0.45\textwidth]{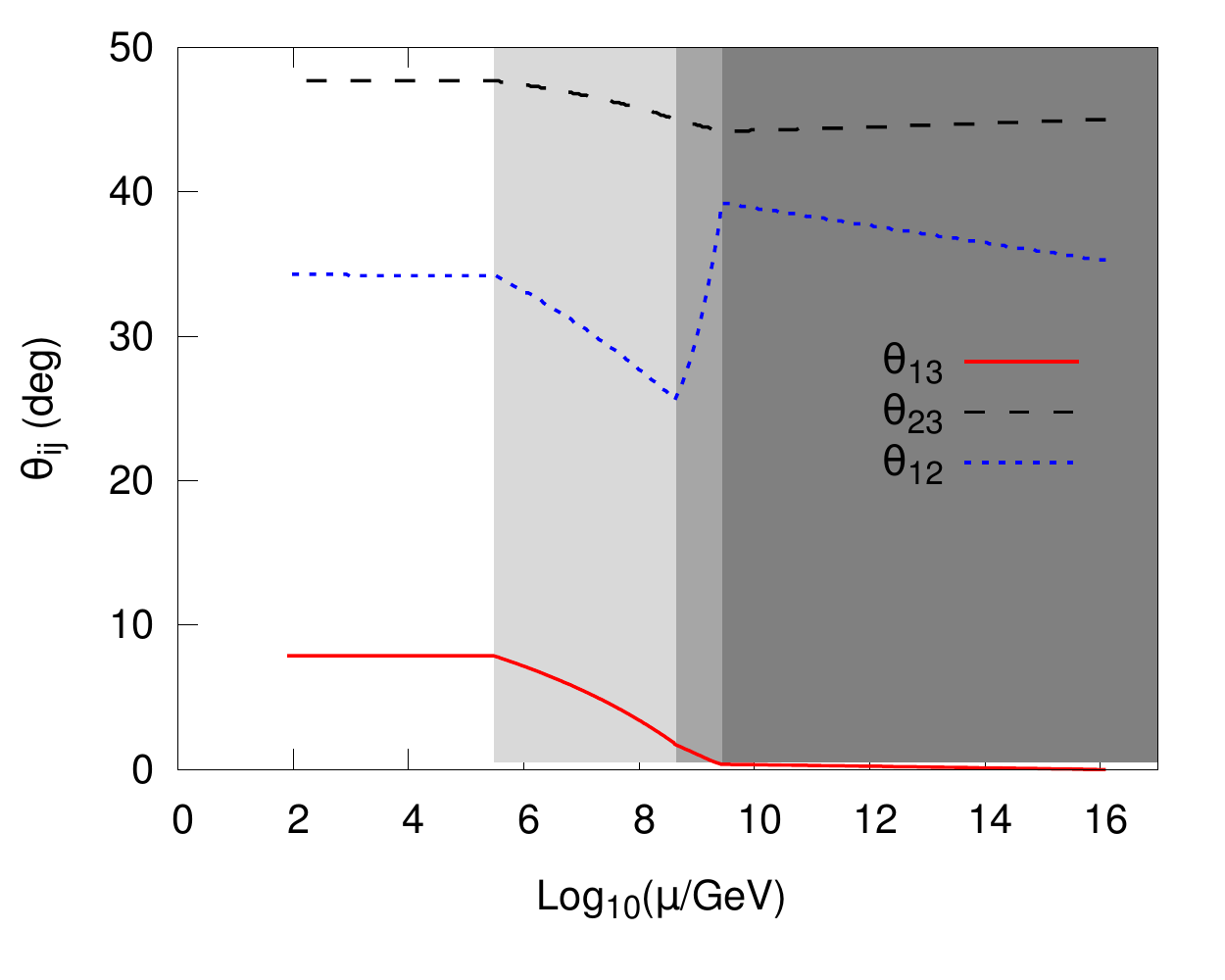}
\includegraphics[width=0.45\textwidth]{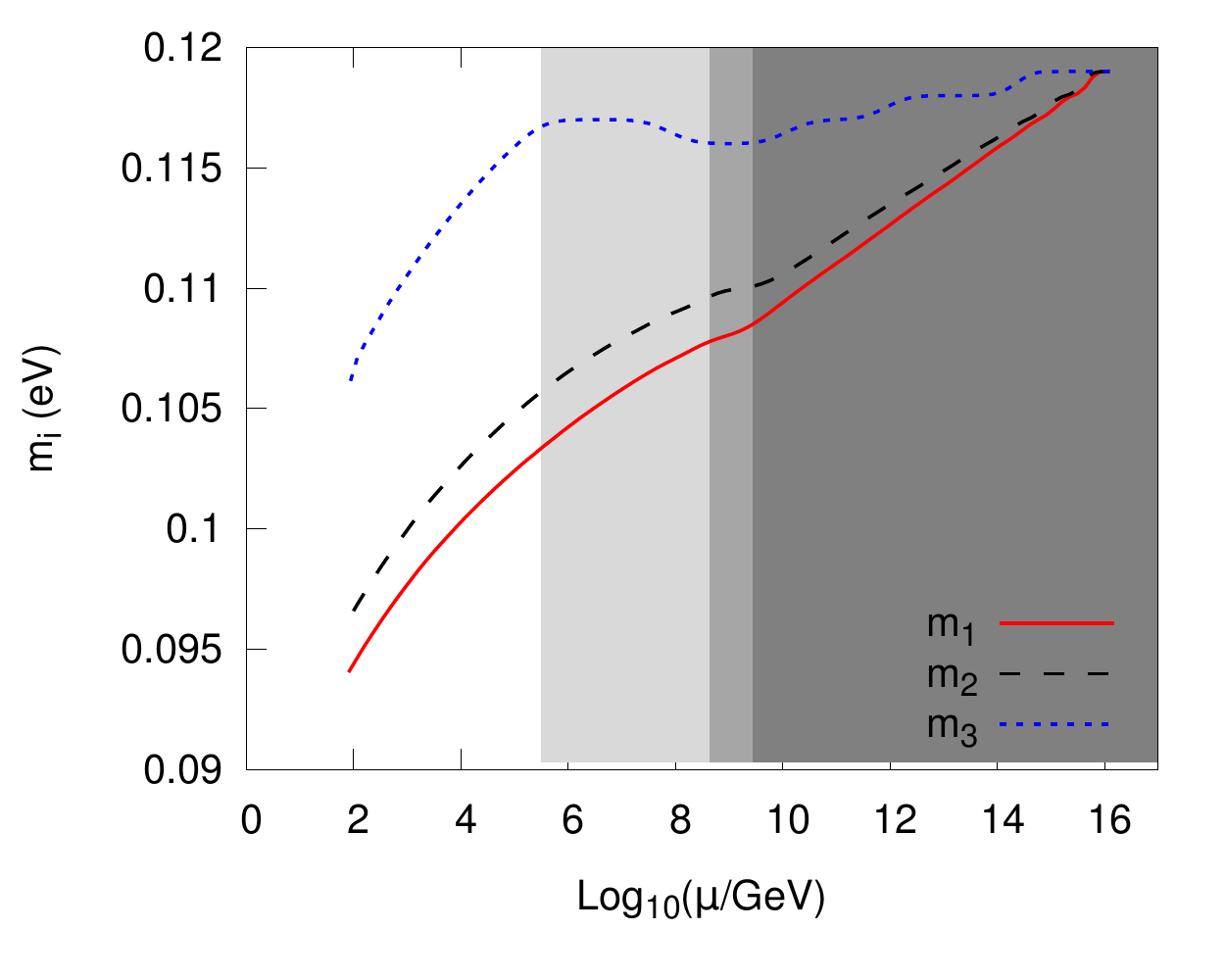} \\
\includegraphics[width=0.5\textwidth]{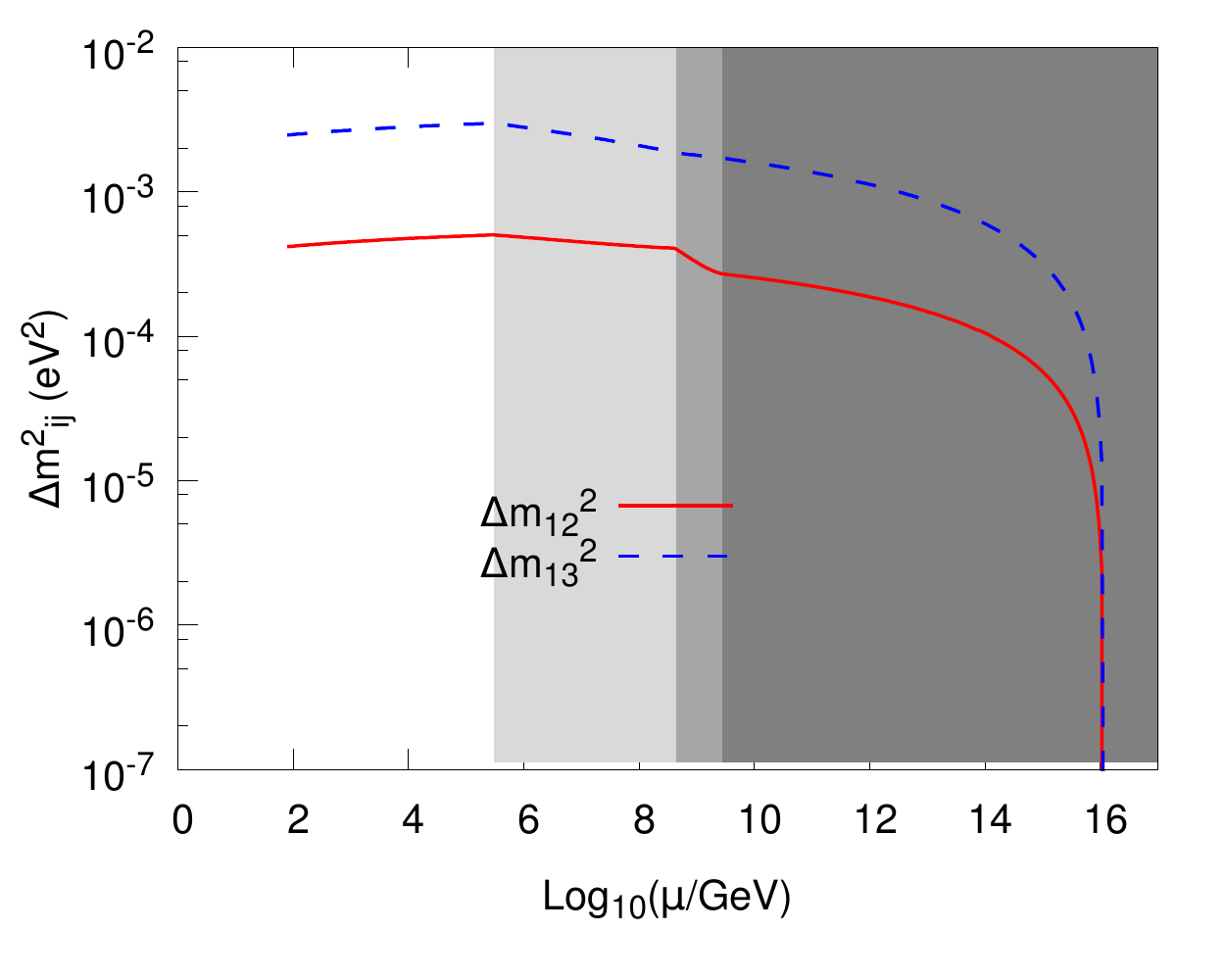}
\end{center}
 \begin{center}
\caption{The RGE of the mixing angles, masses and mass squared
differences between the GUT scale $\Lambda_{g}$ and the electroweak scale $\Lambda_{ew}$ in the SM.
The initial values of the parameters are given in the second column of Table II.
The boundaries of three grey shaded areas, $i.e.$ dark, medium and light denote the
points when heavy right handed singlets $M_3$, $M_2$ and $M_1$ are integrated out respectively.}
\end{center}
\end{figure}

Study of radiative corrections to $S_3$ symmetric neutrino mass matrix can be divided into three regions
that are governed by different RGE equations.
The first region is above the highest seesaw scale $M_3$ to $\Lambda_g$, where there can be considerable contribution of $Y_{\nu}$.
The second region is the region between the three seesaw scales and the third is below the lowest seesaw scale $M_1$,
where all heavy fields are decoupled.
The solar mixing angle $\theta_{12}$ can have large RGE corrections as the running is enhanced by the factor proportional to
$\frac{m^2}{\Delta m_{12}^2}$ at the leading order, which can be large for degenerate spectrum.
The RGE is comparatively small for other two mixing angles $\theta_{23}$ and $\theta_{13}$,
where the RGE is proportional to $\frac{m^2}{\Delta m_{13}^2}$.
However, for the degenerate neutrino mass spectrum there can be considerable corrections for these mixing angles, too.
Below the seesaw scale the RGE corrections to the mixing angles in the SM are negligible
as they get contributions only from $Y_{\tau}$.
In Fig 1, we show the RGE corrections to the mixing angles and masses in the SM for the set of input parameters given
in second column of Table II. Fig. 1 shows that below $M_1$
scale there is no significant corrections to the mixing angles.
Below the lowest seesaw scale the running of the mass eigenvalues is significant even in the SM for
degenerate as well as hierarchical neutrinos \cite{Antush2}
due to the factor $\alpha$ given in Eq. (\ref{alphasm}), which is much larger than $Y_{\tau}^2$.
The running of masses is given by a common scaling of the mass
eigenvalues \cite{chankowski}. Clearly, the RGE of each mass eigenvalue is proportional to the mass eigenvalue itself.
The running of masses in Fig. 1 can be seen to start from the degenerate values of masses at $\Lambda_{g}$
and there are significant corrections to the masses below $M_1$.
Earlier analyses \cite{Casas} studied the successful generation of mass squared differences
and mixing angles for degenerate neutrinos in the SM.
In their analysis it is shown that the generation of mass squared differences is very
sensitive to the value of $\sin^2\theta_{12}$, which should be greater than
0.99 to fit the mass squared differences simultaneously with the consistent angles.
This limit is, however, completely ruled out by the current oscillation data.
The degeneracy of three mass eigenvalues is lifted by the RGE running from $\Lambda_{g}$ to $M_3$.
Potentially significant breaking of neutrino mass degeneracy is provided by the RGE effects.
The seesaw threshold effects in addition increase the mass splitting
between the masses $m_2$ and $m_3$ required to fit the masses with the current data in terms of mass squared differences.

Running between and above the seesaw scales is modified by the contribution of Yukawa couplings $Y_{\nu}$.
The contribution of $Y_{\nu}$ in the RGE can result
all the three mixing angles in their current allowed ranges at the EW scale in the SM.
Thus, there are significant corrections to mixing angles even in the SM in the presence of the seesaw threshold effects
when there is exactly the degenerate mass spectrum to begin with.
As can be seen from Fig. 1,
the mixing angles at the GUT scale are $\theta_{23}=45^{\circ}$, $\theta_{12}=35.3^{\circ}$ and $\theta_{13}=0^{\circ}$.
For the set of parameters given in the second column of Table II we get
the mixing angles in the allowed range at the electroweak scale in the SM.
$\theta_{23}$ is found to have values below maximality.
The presence of seesaw threshold corrections and thus $Y_{\nu}$ in the RGE
equations make this possible even in the SM, as can be seen in Fig. 1.
The grey shaded area in Fig. 1 and Fig. 2 illustrates the ranges of effective
theories that emerge when we integrate out heavy right handed singlets.
At each seesaw scales, $i.e.$ $M_1$, $M_2$ and $M_3$, one heavy singlet
is integrated out and thus (n-1)$\times$3 sub-matrix of $Y_{\nu}$ remains.
Therefore, the running behavior between these scales can be different
from running behavior below or above these scales.
Between these scales the neutrino mass matrix comprises of two terms $\kappa_{(n)}$
and 2$Y_{\nu(n)}^T M_{R(n)}^{-1}Y_{\nu(n)}$, as given in Eq. (\ref {Mnuthr}).
It is shown in \cite{Antush2} that in the SM these two terms between the
thresholds are quite different which can give dominant contribution to
the running of mixing angles in this region. Both $\theta_{12}$ and $\theta_{13}$
in Fig. 1 get large corrections between three
seesaw scales. $\theta_{23}$ gets the deviation of $\approx$2.7$^\circ$ in the upper direction.

Three neutrino mixing angles can be generated in their current allowed ranges starting
from $S_3$ symmetric neutrino mass matrix at the $\Lambda_{g}$
in the SM as there can be large corrections when we consider the seesaw threshold effects.
In the SM, atmospheric mass squared difference $\Delta m^2_{13}$ is generated
within the current oscillation data limit ($\approx2.48\times10^{-3}$)
starting from the vanishing value at the $\Lambda_g$ since all masses are degenerate, as seen from Fig. 1.
The solar mass squared difference $\Delta m^2_{12}$ of the order of
$\approx$ 10$^{-4}$eV$^2$ is simultaneously generated at the $\Lambda_{ew}$, as shown in Fig.1 which
is larger than its present value. The byproduct of this analysis is the masses of right-handed
neutrinos that are determined from Eqs. (\ref{MR}) and (\ref{MR2}) and are not free parameters.
The values of effective Majorana mass $|M_{ee}|$ and Jarlskog rephasing invariant ($J$) at the $\Lambda_{ew}$
are also calculated for particular set of parameters given in Table II.
We conclude that in the SM the RGE in addition to seesaw threshold effects cannot simultaneously
produce the solar mass squared difference in its allowed range at low scale along with the other neutrino oscillation parameters.

\subsection{Radiative Threshold corrections in the MSSM}

\begin{figure}[b]
\begin{center}
\label{figmssm}
\includegraphics[width=0.45\textwidth]{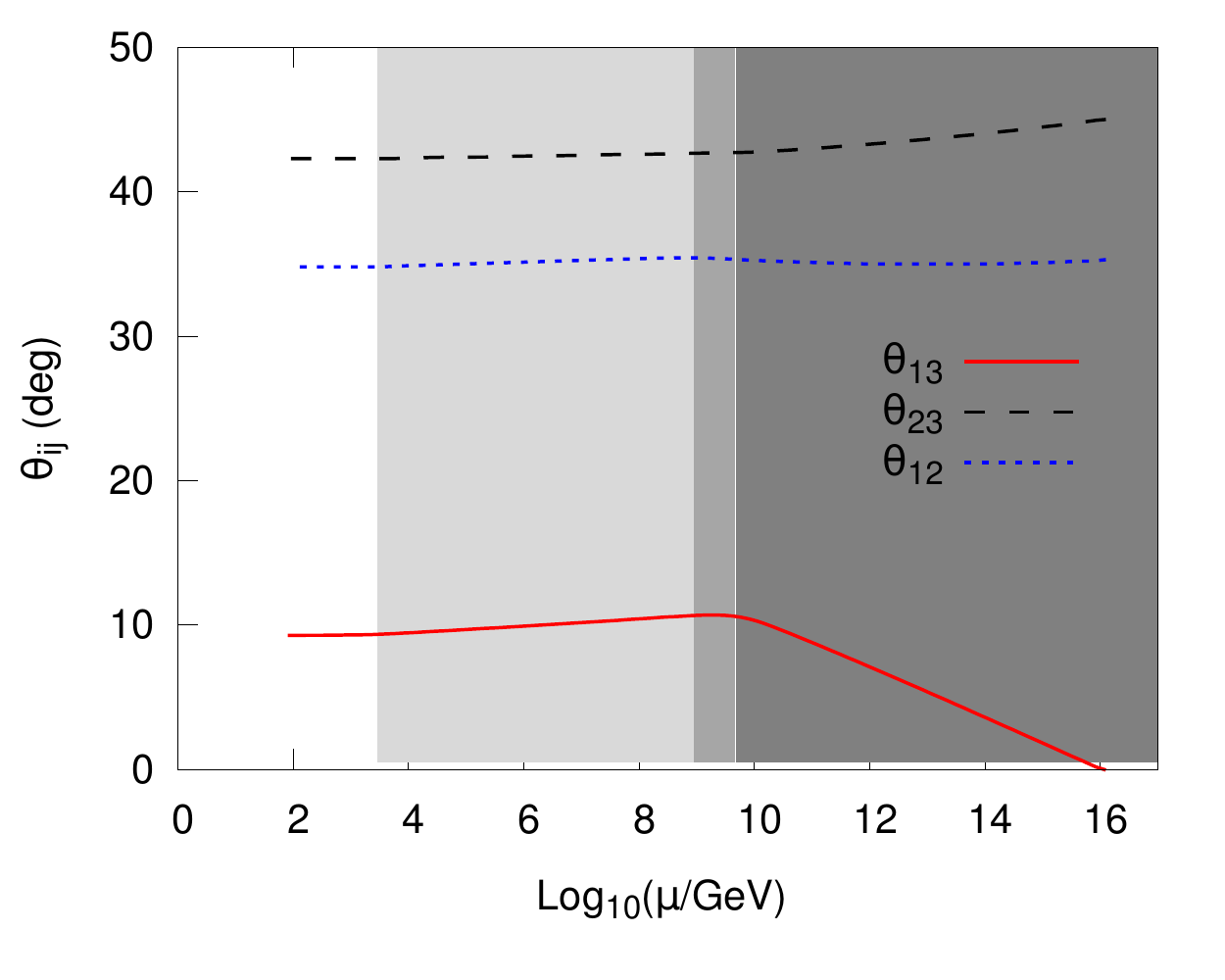}
\includegraphics[width=0.45\textwidth]{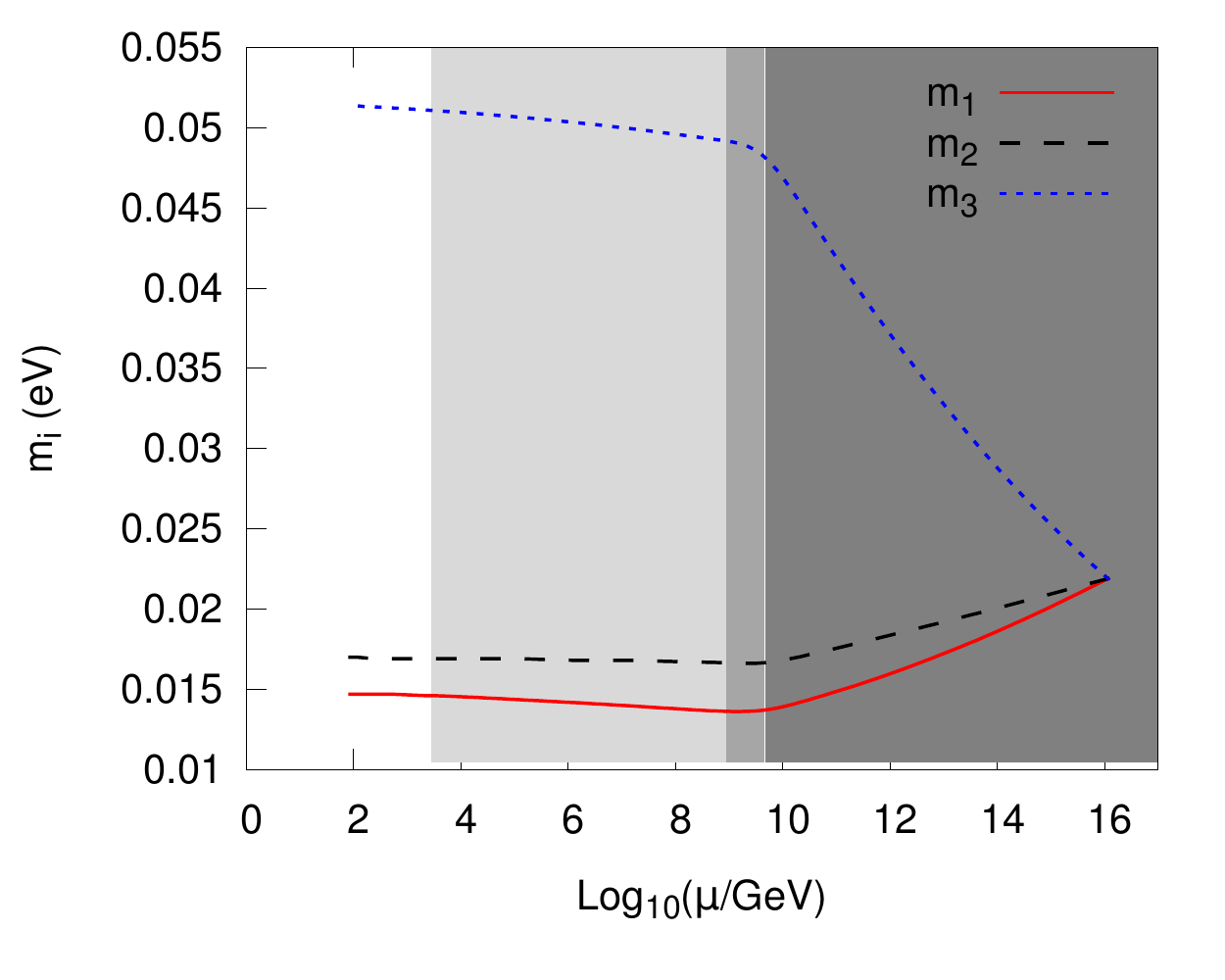} \\
\includegraphics[width=0.5\textwidth]{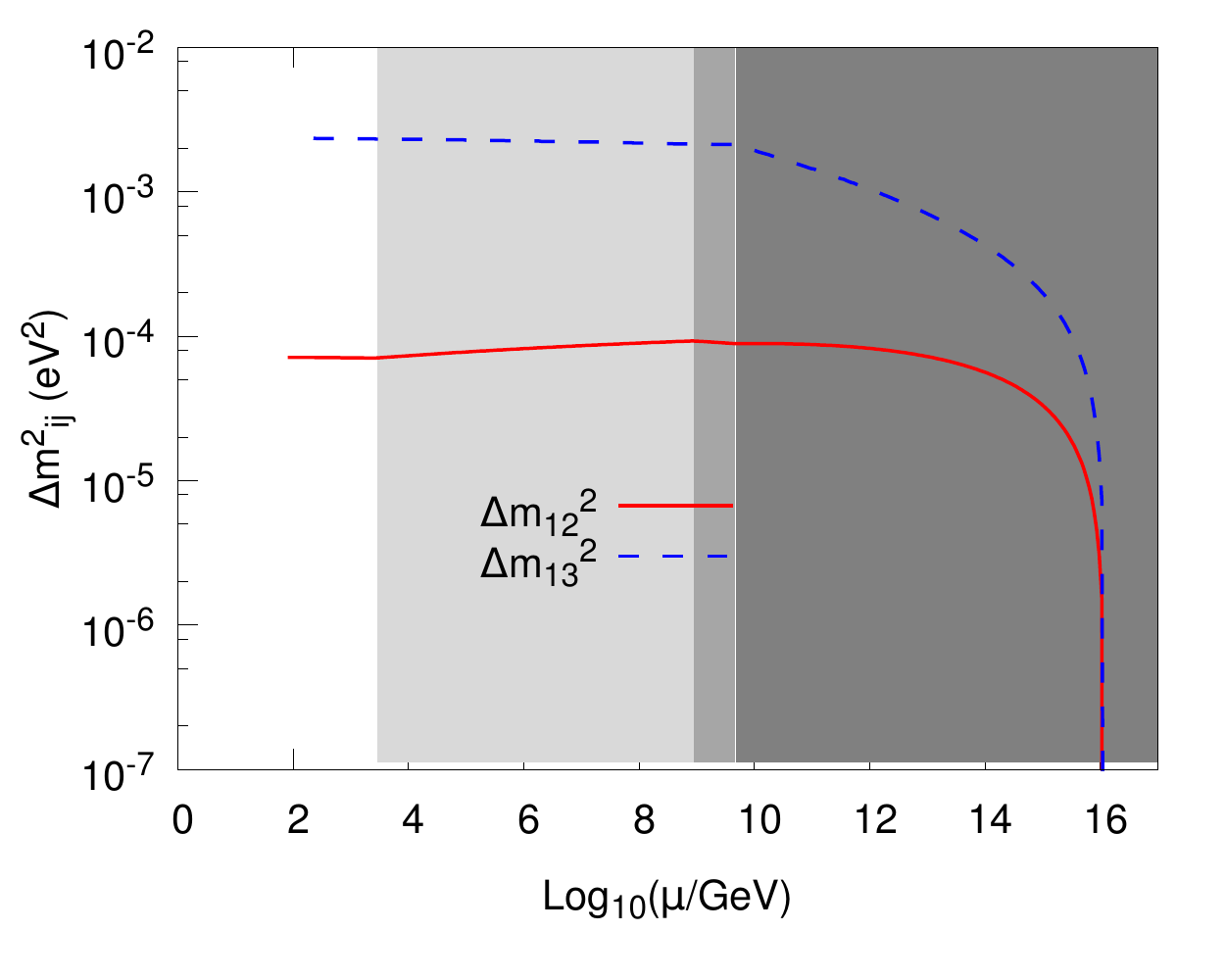}

\end{center}
\begin{center}
\caption{The RGE of the mixing angles, masses and  mass squared differences between the GUT scale $\Lambda_{g}$
and the electroweak scale $\Lambda_{ew}$ in the MSSM with tan$\beta$=55. The initial values of
the parameters are given in the third column of Table II.
The boundaries of three grey shaded areas, $i.e.$ dark, medium and light denote the
points when heavy right handed singlets $M_3$, $M_2$ and $M_1$ are integrated out respectively.}
\end{center}
\end{figure}

As stated earlier we divide the radiative corrections to $S_3$ symmetric neutrino mass matrix
into three regions governed by different RGE equations in the MSSM.
In the region below the lightest seesaw scale, for the SM and MSSM with small tan$\beta$, Yukawa coupling $Y_{\tau}$ $\sim$0.01 is small,
and thus there cannot be large RGE corrections in these cases.
For large tan$\beta$ these corrections can be larger due to the presence of factor $Y_{\tau}^2(1+\tan^2\beta)$
in the MSSM. We show the RGE of mixing angles and masses in the MSSM with tan$\beta$=55
in Fig. 2 for the set of input parameters given in third column of Table II.
In the region above the energy scale $M_1$, we get contributions from another Yukawa coupling $Y_\nu$ which brings in more
free parameters in the analysis. In the region for the MSSM with large tan$\beta$
the presence of seesaw threshold effects can enhance the RGE of the mixing angles significantly.
As can be seen from Fig. 2, we can have all three mixing angles simultaneously in the current limit at the EW scale starting
from $S_3$ symmetric neutrino mass matrix at the $\Lambda_g$. Fig. 2 shows that there are
large corrections to $\theta_{13}$ ($\sim$9.85$^\circ$) between the $\Lambda_g$ and $M_3$
scale due to the presence of $Y_{\nu}$.
As mentioned earlier, the running of neutrino mass matrix between the seesaw thresholds
gets contributions from two terms $\kappa_{(n)}$ and 2$Y_{\nu(n)}^T M_{R(n)}^{-1}Y_{\nu(n)}$ given in
Eq. (\ref{Mnuthr}). As shown in the SM, these two terms are different and thus results
in the enhanced running between the different energy regimes. On the other hand, in the MSSM,
as can be seen from Fig. 2, there are not much deviations in the mixing angles between the energy scales.
It is because in the MSSM the two contributions are almost identical and thus cancel each other resulting
in minimum deviation in those regions. The only significant corrections occurs in the region above the highest seesaw scale $M_3$
due to relatively large $Y_\nu$.

The mixing angle $\theta_{12}$ does not have much corrections and $\theta_{23}$ receives the correction of
2.5$^\circ$ in the upper direction and is thus above maximal.
The running of masses in the MSSM (Fig. 2) is much larger than the SM due to
the presence of tan$\beta$ which in our case is large.
The dominant effect, however, is the corrections in the range $M_3 \leq \mu \leq \Lambda_{g}$
where the flavor dependent terms ($Y_l$ and $Y_{\nu}$) can be large.
The interesting dependence of $\alpha_{\nu}$ (MSSM) and tan$\beta$ on the
running contributions of flavor dependent terms
is given in \cite{Antush2}. For large tan$\beta$ the contribution of $Y_e$ and $Y_{\nu}$ become important.
We also show the radiative corrections to the solar and atmospheric mass
squared difference from the $\Lambda_g$ to the $\Lambda_{ew}$ for degenerate masses at the GUT scale in the MSSM with tan$\beta$=55.
To begin with both the mass squared differences are zero at the GUT scale.
From the mass squared differences shown in Fig 2, we see that the RGE in combination with seesaw threshold corrections can result
both mass squared differences in their current 3$\sigma$ ranges at the low scale.
The hierarchy parameters $y_{\nu}$, $r_1$ and $r_2$ though arbitrary but are found to be $<$ $\mathcal{O}(1)$.
For given set of input parameters in Table II, the value of effective Majorana mass $|M_{ee}|$ is $\approx$10$^{-2}$eV and
that of $J$ is $\approx$3.87 $\times$ 10$^{-2}$.
Thus, we find that it is possible to simultaneously obtain the neutrino oscillation parameters at the  electroweak scale for $S_3$
mass matrix at the $\Lambda_{g}$ in the MSSM with large tan$\beta$.

\section{Conclusions}

We studied the RGE corrections to the $S_3$ symmetric neutrino mass matrix
in the presence of seesaw threshold corrections both in the SM and MSSM.
In the absence of seesaw threshold effects there are negligible corrections to the mixing angles in the SM and MSSM with low tan$\beta$.
However, large corrections are possible in neutrino parameters once the seesaw threshold effects are taken into consideration both in the SM and MSSM.
Above all seesaw scales there is the contribution of the Yukawa coupling, $Y_{\nu}$ in
addition to $Y_{l}$ and thus the RGE depends on more parameters than below the seesaw
scales. Thus, we can say the corrections depend on the form of $Y_{\nu}$ which has free parameters.
In the SM we found that the mixing angles can be obtained in their current 3$\sigma$ range at the  electroweak scale when we
begin with $S_3$ neutrino mass matrix at the GUT scale $\Lambda_g$.
The significant running occurs between and above the seesaw threshold scales.
Below the lowest seesaw scale there are no significant corrections as the only contribution comes from $Y_{\tau}$ which is small.
However, in this case of exactly equal magnitude of mass eigenvalues, the two mass squared difference
are not simultaneously generated in the current range at the electroweak scale.
The solar mass squared difference is found to be large $(\mathcal{O}(10^{-4}))$ in comparison to its allowed value at the  electroweak scale.

There can be large radiative corrections in the MSSM with tan$\beta$=55 when threshold effects are taken into consideration.
The large corrections to the mixing angles occur at the scale above the seesaw threshold
where the Yukawa coupling, $Y_{\nu}$, is present and has large free parameters which can enhance running for large tan$\beta$.
All the mixing angles and the mass squared differences are obtained in their current 3$\sigma$ range at the low scale in this case.
The free input parameters $y_{\nu}$, $r_1$ and $r_2$
are expected to be small $< \mathcal{O}(1)$.
Thus, for the SM it is not possible to radiatively generate the solar mass squared difference at the  electroweak scale in its current range.
In the MSSM with large tan$\beta$ we can generate all the masses and mixing angles in the current allowed range at the  electroweak scale, starting from exactly
degenerate mass spectrum at the GUT scale.

\section{Acknowledgements}

The work of C. S. K and S. G. is supported by the National Research Foundation of Korea (NRF) grant funded by
Korea government of the Ministry of Education, Science and Technology (MEST) (Grant No. 2011-0017430) and (Grant No. 2011-0020333).

\end{document}